\documentclass{article}
\usepackage{spconf,amsmath,graphicx}
\usepackage{cite}
\usepackage{url}
\usepackage{graphicx}
\usepackage{amssymb,amsmath,bm}
\usepackage{textcomp}
\usepackage{booktabs}
\usepackage{hyperref}
\usepackage{verbatim}
\usepackage{multirow}
\usepackage[table,xcdraw]{xcolor}
\usepackage{array}
\usepackage{tabularx}
\usepackage{bbding}
\usepackage{color} 
\newcolumntype{Y}{>{\centering\arraybackslash}X}
\newcommand{\norm}[1]{\left\lVert#1\right\rVert}
\title{AudioDec: An Open-source Streaming High-fidelity Neural Audio Codec}
\name{Yi-Chiao Wu, Israel D. Gebru, Dejan Markovi\'{c}, Alexander Richard}
\address{Meta Reality Labs Research, USA}
\begin{document}
\ninept
\maketitle
\thispagestyle{plain}
\pagestyle{plain}
\begin{abstract}
A good audio codec for live applications such as telecommunication is characterized by three key properties: (1) compression, i.e.\ the bitrate that is required to transmit the signal should be as low as possible; (2) latency, i.e.\ encoding and decoding the signal needs to be fast enough to enable communication without or with only minimal noticeable delay; and (3) reconstruction quality of the signal. In this work, we propose an open-source, streamable, and real-time neural audio codec that achieves strong performance along all three axes: it can reconstruct highly natural sounding 48~kHz speech signals while operating at only 12~kbps and running with less than 6~ms (GPU)/10~ms (CPU) latency. An efficient training paradigm is also demonstrated for developing such neural audio codecs for real-world scenarios. Both objective and subjective evaluations using the VCTK corpus are provided. To sum up, AudioDec is a well-developed plug-and-play benchmark for audio codec applications.
\end{abstract}
\section{Introduction}
\label{sec:intro}
\vspace{-\baselineskip}
An audio codec is a technique to compress audio signals into codes and reconstruct the audio signals on the basis of the codes. A typical audio codec system is composed of encoder, quantizer, and decoder modules. The bitrate of the quantized codes is usually much lower than that of the input audio signals, so the codes are suitable for transmissions or storage. Audio codec techniques have been applied to a variety of real-world applications such as secure communication~\cite{codec1976, codec1994-1}, low-cost mobile and internet communications~\cite{codec1986, codec1994-2}, and live videos and music streamings ~\cite{codec1996}.

Most of the conventional parametric codecs~\cite{codec1970, celp, lpc} were built according to in-domain knowledge of psychoacoustics, human speech production systems, and traditional digital signal processing. Although the bitrate is low, the quality is also low because of bandwidth limitations. Modern parametric audio codecs~\cite{amrwb, opus, evs} usually achieve acceptable reconstruction quality with 16~kHz or higher sampling rate audio signals as a consequence of the advanced transmission techniques. However, the ad~hoc designs and limited modeling capacity of these codecs still result in a significant quality gap between natural and reconstructed audio signals.

Recently, the rapid developments of neural networks (NNs) provide advanced modeling capacity, so many neural codecs have been proposed in past years. The first category is the hybrid codec, which replaces or integrates part of the parametric codecs’ modules with NNs to improve the compression or reconstruction performances. For example,  Krishnamurthy et al.~\cite{hycodec1990} proposed a neural vector quantizer for speech and image coding. Wu et al.~\cite{hycodec1994}. proposed a neural predictive speech coder with coded excitation inputs. NN-based encoder-decoder structures also have been proposed for different handcraft acoustic features such as spectrograms~\cite{hycodec2010}, phonological features, and pitches~\cite{hycodec2016}. However, these codecs still require lots of ad~hoc designs for speech signals.

The second category is the vocoder-based codec, which directly reconstructs audio signals using neural vocoders conditioned on the codes from other parametric coders~\cite{wavenetcodec} or quantized acoustic features~\cite{samplernncodec, lpcnetcodec, melgancodec}. However, the performance and bitrate of vocoder-based codecs are still bonded with the upstream handcraft encoding. To theoretically achieve global optimations and flexible bitrates, end-to-end autoencoders (E2E AEs) with raw waveform I/O~\cite{ aecodec1990, aecodec2018, vqvae, vqvae2019, cmrl, soundstream, harpnet, csvq} recently have been investigated. Although these E2E codecs usually attain high-fidelity speech generations, the open-source benchmark is unavailable, the efficient training paradigm is unclear, the comparison to vocoder-based methods is absent, and the system for different applications is inflexible.

To tackle these issues, we present an open-source E2E neural codec, AudioDec\footnote{\label{repo}\url{https://github.com/facebookresearch/AudioDec}}, with an efficient training paradigm in this paper. The modularized architecture provides the flexibility of developing systems for different applications. Specifically, since both the encoder and decoder are easily replaceable, we can separately develop several specific encoders and decoders for any applications and easily integrate or switch among them for different real-world scenarios such as binaural rendering~\cite{binauralss}. One of the state-of-the-art neural vocoders, HiFi-GAN~\cite{hifigan}, is integrated into our codec, and we argue that the combination of a separate powerful vocoder and a well-trained encoder will attain the highest quality. For practicality, we adopt a group convolution mechanism to make the streamable network run in real-time with low latency on both GPU and CPU. Both objective and subjective experiments are conducted, and the comparisons to the vocoder-based mode are also presented. The experimental results show the effectiveness of the proposed codec to generate high-fidelity 48~kHz speech and give more insight into efficiently building a practical neural codec.

\section{Baseline neural audio codec}
\label{sec:baseline}
\vspace{-\baselineskip}
Since AudioDec adopts an E2E AE-based architecture with the SoundStram backbone~\cite{soundstream}, the E2E AE-based audio codec foundations and the baseline SoundStream are introduced in this chapter.

\subsection{End-to-end Autoencoder-based Audio Codec}
A typical E2E AE-based codec consists of encoder, projector, quantizer, and decoder modules. In the encoding stage, raw waveform signals are encoded into representations with a much lower temporal resolution and then projected to the designed multidimensional space. The projected representations are further quantized into codes for transmission or storage. In the decoding stage, the codes are first transferred to the representations by a lookup process, and then the decoder reconstructs the raw waveform based on the representations. Morishima et al.~\cite{aecodec1990} proposed the first NN-based AE codec with raw waveform I/O, but the quantizer applied to the bottleneck features is not jointly trained.  

To jointly optimize all modules, many E2E AEs adopting vector quantization (VQ)~\cite{vq1985} have been proposed to tackle the gradient of the VQ. For example, the neural codecs with softmax quantization~\cite{aecodec2018}, straight-through gradient, exponential moving average (EMA)~\cite{vqvae, vqvae2019}, and Gumble-softmax~\cite{gumbelvqvae} have been recently proposed and work well for gradient propagation. In addition, the bitrate is directly related to the VQ codebook size, but training the model with a huge plain codebook is impractical. Therefore,  scalable codecs decomposing the fine-coarse structures of the encoded latent codes using residual VQ~\cite{soundstream}, multi-scale VQ~\cite{harpnet},  and cross-scale VQ~\cite{csvq} or the output waveforms using cross-module residual learning~\cite{cmrl} are introduced for tractable hierarchical codebooks.

\subsection{SoundStream Audio Codec}
SoundStream is an E2E AE-based neural codec adopting the residual VQ mechanism. For 24~kHz audio signal coding, SoundStream is comparative to the modern state-of-the-art parametric codecs, Opus~\cite{opus} (12~kbps) and EVS~\cite{evs} (9.6~kbps),  with only 3~kbps while a single SoundStream model can work on different bitrate from 3~kbps to 18~kbps. To meet the streamable and real-time requirements, SoundStream adopts a fully causal convolution architecture. The causality makes the network encode/decode audio signals based on only previous samples, so the whole process can run in a continuous segmental manner. The convolution network takes advantage of parallel computations for efficient encoding/decoding. 

Both metric and adversarial losses are adopted to train the model. Specifically, given the input signal $\boldsymbol{x}$ and the output signal $\hat{\boldsymbol{x}}$,  the adopted metric mel spectral loss $L_{\mathrm{mel}}$ is formulated as
\begin{align}
L_{\mathrm{mel}}(\boldsymbol{x}, \hat{\boldsymbol{x}})
=\mathbb{E}\left[\norm{\mathrm{mel}(\boldsymbol{x})-\mathrm{mel}(\hat{\boldsymbol{x}})}_{1}\right],
\label{eq:melloss}
\end{align}
where $\mathrm{mel}()$ denotes the mel spectrogram extraction. Two types of fully convolutional discriminators are utilized, and the main difference is that the short-time Fourier transform discriminator (STFTD) takes a complex spectrogram as the input while the multi-scale discriminators (MSDs)~\cite{melgan} take a waveform as the input. All discriminators adopt the hinge loss over the discriminator output logits. Given a discriminator $D$ and a generator $G$, the adversarial discriminator loss $ L_{\mathrm{D}}$ is defined as
\begin{align}
\label{eq:dloss}
L_{\mathrm{D}}=\mathbb{E}_{\boldsymbol{x}}\left[\mathrm{max}(0, 1-D(\boldsymbol{x})) +\mathrm{max}(0, 1+D(G(\boldsymbol{x})))\right], 
\end{align}
and the adversarial generator loss $L_{\mathrm{adv}}$ is defined as
\begin{align}
\label{eq:gloss}
L_{\mathrm{adv}}=\mathbb{E}_{\boldsymbol{x}}\left[\mathrm{max}(0, 1-D(G(\boldsymbol{x})))\right].
\end{align}
Moreover, the feature matching loss~\cite{melgan} $L_{\mathrm{fm}}$ is applied to the feature maps of all discriminators, and the EMA~\cite{vqvae} loss $L_{\mathrm{vq}}$ is applied to the VQ codebook. Therefore, the overall generator loss is
\begin{align}
\label{eq:ssloss}
L_{\mathrm{G}}=L_{\mathrm{adv}}+\lambda_{\mathrm{fm}} L_{\mathrm{fm}}+\lambda_{\mathrm{mel}} L_{\mathrm{mel}}+\lambda_{\mathrm{vq}} L_{\mathrm{vq}},
\end{align}
where $\lambda_{\mathrm{fm}}$, $\lambda_{\mathrm{mel}}$, and $\lambda_{\mathrm{vq}}$ are the weights.

Although SoundStream achieves high-quality audio reconstruction with a low bitrate, the training efficiency and the model flexibility can be improved. Specifically, in contrast to the very lightweight generator, a generative adversarial network (GAN)-based model usually requires multiple deep discriminators to achieve high-fidelity generation~\cite{ganvocoder}, and training these discriminators is time-consuming. However, the GAN training is mostly related to improving the waveform details, high-frequency components, and phase synchronization while modeling the low-frequency component can be learned solely with the metric losses. As a result, directly training the model with both metric and adversarial losses from scratch is inefficient. In addition, training a specific SoundStream model from scratch for each scenario is required, but a practical system should be flexibly adjusted for different scenarios such as switching between mono and binaural outputs according to the user's hardware. 

\section{Proposed neural audio codec}
\label{sec:proposed}
\vspace{-\baselineskip}
To improve the training efficiency, model flexibility, and audio quality of SoundStream, we propose an efficient training paradigm and a modularized architecture and adopt the HiFi-GAN-based multi-period discriminator for developing the AudioDec codec. Moreover, we also apply the group convolution mechanism to the HiFi-GAN vocoder to make it run in real-time on a CPU with 4 threads.

\begin{figure}[t]
\centering
\centerline{\includegraphics[width=0.95\columnwidth]{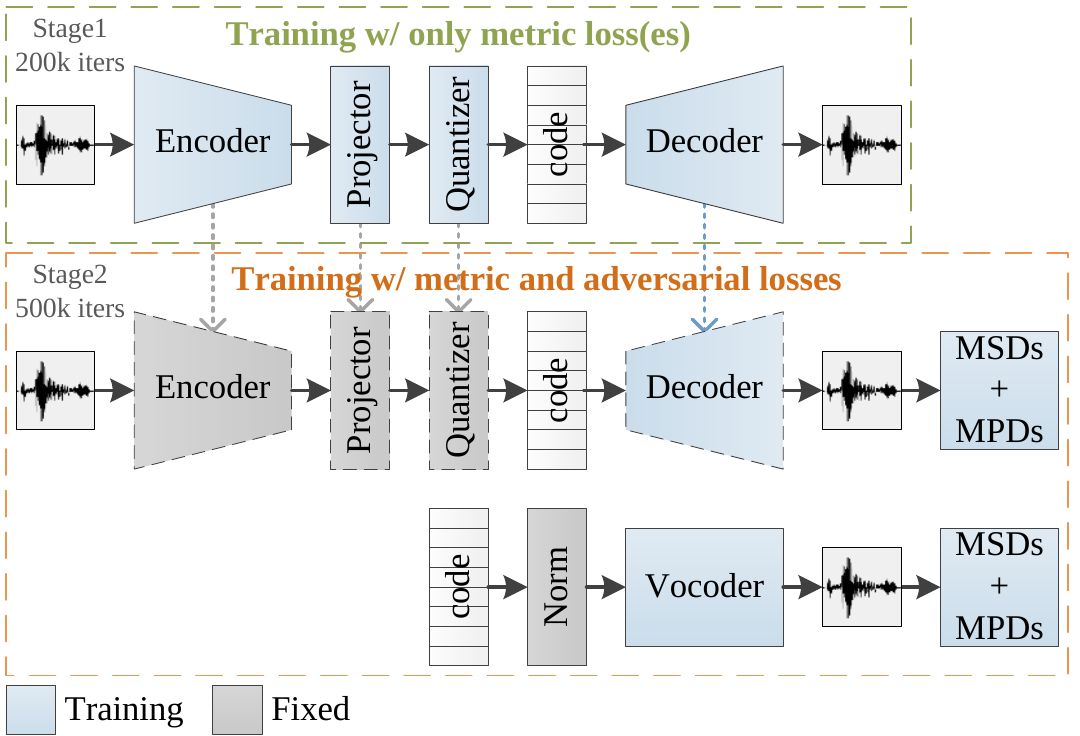}}
\caption{Training paradigm of AudioDec.}
\label{fig:audiodec}
\end{figure}

\subsection{Efficient Training Paradigm}
According to recent speech research, we know that although standard spectral features such as mel spectrogram already include the most high-level information of speech such as contents and speaker identity, the phase information is crucial for generating high-fidelity speech. In addition, for a codec, the codes are expected to contain only essential information for low-bitrate transmissions, and the decoder should be powerful enough to reconstruct high-fidelity waveforms. As a result, we propose an efficient training paradigm to first train both the encoder and decoder with only the metric loss, which makes the training converge fast and stable. Then the discriminators are jointly trained with only the decoder to tackle the details and phase synchronizations of the reconstruction waveforms.

As shown in Fig.~\ref{fig:audiodec}, given the encoder (including the projector and quantizer)  parameters $\theta$ and the decoder parameters $\phi$, the AudioDec generator is first trained using eq.~\ref{eq:melloss} for the first 200k iterations (stage1). For the following 500k iterations (stage2), the whole model is trained with the updated mel spectral loss:
\begin{align}
L_{\mathrm{mel}'}(\boldsymbol{x}, \hat{\boldsymbol{x}})
=\mathbb{E}\left[\norm{\mathrm{mel}(\boldsymbol{x})-\mathrm{mel}(f_{\phi}(\mathrm{sg}[f_{\theta}(\boldsymbol{x})]))}_{1}\right],
\label{eq:umelloss}
\end{align}
where $\mathrm{sg}[]$ denotes the stop gradient operator, the updated adversarial discriminator loss: 
\begin{align}
\label{eq:udloss}
L_{\mathrm{D'}}=\mathbb{E}_{\boldsymbol{x}}\left[(1-D(\boldsymbol{x}))^{2} +D(f_{\phi}(\mathrm{sg}[f_{\theta}(\boldsymbol{x})]))^{2}\right], 
\end{align}
and the updated adversarial generator loss:
\begin{align}
\label{eq:ugloss}
L_{\mathrm{adv'}}=\mathbb{E}_{\boldsymbol{x}}\left[(1-D(f_{\phi}(\mathrm{sg}[f_{\theta}(\boldsymbol{x})])))^{2}\right].
\end{align}
The least squares-GAN is adopted to improve the training stability.

\subsection{Modularized Architecture}
Since the encoder and decoder may be rapidly replaced for different scenarios such as denoising and binaural rendering, modularizing each component of a codec is essential for simultaneously developing encoders and decoders. Inspired by the pretrain mechanism~\cite{gumbelvqvae, binauralss}, the proposed AudioDec model is first trained with a high-quality clean corpus to obtain a standard quantizer and a complete codebook. By fixing the quantizer and codebook, we can easily develop arbitrarily encoders and decoders for any new scenarios.

Furthermore, we find that the symmetric encoder-decoder architecture is essential for the training stability of an AE. However, the waveform decoder is usually expected to be more powerful to cover all details of the high-resolution audio signals while the encoder is expected to preserve limited essential information of the input audio signals. That is, asymmetric architecture tends to be unstable, and symmetric powerful architecture is inefficient. Therefore, the modularized architecture provides a more flexible design that the decoder of a well-trained lightweight AE codec can be easily replaced by powerful vocoders, and the following experiments actually show that running AudioDec in a vocoder-based mode using a separately trained HiFi-GAN vocoder achieves the best performance.

\subsection{HiFi-GAN-based Multi-Period Discriminator}
HiFi-GAN is one of the state-of-the-art neural vocoders achieving very high-fidelity speech generation. The main breakthrough of HiFi-GAN is the effectiveness of the proposed multi-period discriminator (MPD)~\cite{hifigan}. Specifically, different from MSDs~\cite{melgan} working on the original and downsampling signals, MPDs work on the segmental signals with different segment lengths (periods). Compare to MSD capturing long-term dependency, MPD is effective to capture the periodic details. Since STFTD is also more related to long-term dependency, we find that replacing the redundant STFTD with MPDs does improve the audio quality.

\subsection{Low-latency Implementation}
Streaming and real-time are two main factors of low-latency encoding/decoding. The causal convolutions and deconvolutions of AudioDec are implemented using one-side padding for streaming. Non-autoregressive (Non-AR) architecture is adopted to achieve real-time coding by parallel computations. However, because the multi-receptive field fusion (MRF) module of HiFi-GAN is not parallel-computation-friendly, it is difficult to run the vocoder-based AudioDec with the vanilla HiFi-GAN generator in real-time. The different kernel sizes of each MRF hinder fully parallel computations, but adopting the largest kernel sizes for all MRFs theoretically attains the same or even better modeling ability. In this paper, we find that the ad~hoc kernel sizes are unnecessary, so we utilize the group convolution mechanism~\cite{alexnet} to simulate the MRFs with the same kernel size. The proposed group HiFi-GAN generator greatly improves its running time on both GPU and CPU.

\section{Experiments}
\label{sec:typestyle}
\vspace{-\baselineskip}
\subsection{Experimental Setting}
The Valentini dataset~\cite{noisyvctk}, which is derived from the VCTK corpus~\cite{vctk2017}, is adopted for the evaluations. This English corpus includes 84 gender-balance speakers with accents from England, Scotland, and United States for training and two England speakers (female p257 and male p232) for testing. The number of available utterances of each speaker is around 400. The sample rate of all data is 48~kHz.

The baseline SoundStream (SS), two symmetric AudioDec (symAD), one asymmetric AudioDec (asymAD), and three vocoder-based AudioDec (AD) systems are evaluated in this chapter. Specifically, to evaluate the effectiveness of the proposed training paradigm, two AudioDecs with symmetric encoder-decoder were respectively trained with and without fixing the encoder during the training of the decoder and discriminators. To investigate the importance of symmetric structure, an asymmetric AudioDec adopting a powerful HiFi-GAN-like decoder with the lightweight AudioDec encoder was trained following the proposed training paradigm. To compare with the vocoder-based approach, three HiFi-GANs including vanilla MRF with kernel sizes 3, 7, and 11 (v0), three group convolutions with kernel size 11 (v1), and three group convolutions with kernel size 3 (v2) were trained using the global normalized codes extracted from the natural waveforms by the well-trained AudioDec encoder.

The neural network architecture and hyperparameters of SS, symADs, asymAD, and ADs followed the settings in the binaural SS paper~\cite{binauralss}. Compared with SS, the main modifications of the proposed AudioDec models are the training paradigm, modularized architecture for switching between symmetric and vocoder-based modes, replacing STFTD with MPDs, and adopting LS-GAN for training. The HiFi-GAN-based vocoders followed the popular open-source repository\footnote{\url{https://github.com/kan-bayashi/ParallelWaveGAN}}. The number of overall training iterations is 700k. The encoders, quantizers, and codebooks of the models with the proposed training paradigm were fixed after the first 200k training. The HiFi-GAN-based vocoders were trained with 500k iterations to match the overall training iterations. The downsampling rate was set to 300, and each code was represented by eight codebooks with a 1024 book size ($8 \times10$~bits), so the bitrate of each codec is 12.8~kbps for 48~kHz audio coding. Details can be referred to our repository\textsuperscript{\ref{repo}}.

\begin{table}[t]
\caption{Objective evaluations of 48~kHz codecs w/ 12.8~kbps}
\label{tb:objective}
\fontsize{8pt}{9.6pt}
\selectfont
{%
\begin{tabularx}{\columnwidth}{@{}p{1.5cm}YYYYY@{}}
\toprule
& $F_{0}$RMSE (Hz)$\downarrow$ & $U/V$ (\%)$\downarrow$ & MCD (dB)$\downarrow$ & LSD (dB)$\downarrow$ & DNSMOS $\uparrow$ \\ \midrule
Natual     & -      & -     & -      & -      & 3.95 \\
\midrule
SoundStream         & 12.5   & 5.2   & 4.42   & 0.89   & 3.81 \\ \midrule
symAD      & 11.8   & 4.9   & 4.36   & 0.89   & 3.88 \\
symAD*     & 12.9   & 5.3   & 4.57   & 0.89   & 3.86 \\ 
asymAD     & 14.1   & 5.6   & 4.45   & 0.90   & 3.78 \\
\midrule
AudioDec~v0    & 12.0   & 4.9   & \textbf{4.28}   & \textbf{0.88}   & \textbf{3.89} \\
AudioDec~v1    & \textbf{10.7}   & \textbf{4.5}   & 4.29   & \textbf{0.88}   & \textbf{3.89}   \\
AudioDec~v2    & 11.8   & 5.0   & 4.33   & 0.89   & 3.86   \\
\bottomrule
\end{tabularx}%
\\symAD*: symAD w/o fixing the encoder
}
\end{table}

\subsection{Objective Evaluation}
Five objective measurements were adopted, and the results are the averages of all testing utterances. Specifically, speech quality and prosody are two main factors for evaluating codecs. To evaluate the speech prosody, root mean square errors of fundamental frequency ($F_{0}$RMSE) and unvoice/voice ($U/V$) errors were adopted. To evaluate the speech quality, mel-ceptral distortion (MCD), log-spectral distortion (LSD), and non-intrusive speech quality metric Deep Noise Suppression Mean Opinion Score (DNSMOS)~\cite{dnsmos} were adopted. The WORLD vocoder~\cite{world} was utilized to extract $F_0$, $U/V$ flags,  and mel-cepstral coefficients ($mcep$). A public DNSMOS\footnote{\url{https://github.com/microsoft/DNS-Challenge/tree/master/DNSMOS}}  model was adopted. The inputs of the DNSMOS model were downsampled to 16~kHz to match the model, and the SIG\_raw scores (for evaluating clean speech) are reported.

As shown in Table~\ref{tb:objective}, the performance differences between symAD and SS show the effectiveness of the MPDs. The even worse results of the symAD w/o the proposed training paradigm (symAD*) demonstrate that the GAN training is more related to improving the decoder. The worst results of asymAD yield the difficulties of training an asymmetric AE. Furthermore, the proposed vocoder-based codecs (AudioDec v*) achieve the best performance of all measurements, and the group convolution networks work as well as the MRF network showing that the ad~hoc kernel sizes are unnecessary. In conclusion, training a powerful vocoder with the globally normalized codes extracted from a well-trained audio encoder is a efficient and trackable way to build a neural codec.

\begin{table}[t]
\caption{Mean Opinion Scores of 48~kHz codecs w/ 12.8~kbps}
\label{tb:subjective}
\fontsize{8pt}{9.6pt}
\selectfont
{%
\begin{tabularx}{\columnwidth}{@{}YYYYYYY@{}}
\toprule
Natural & \multicolumn{2}{c}{SoundStream} & symAD  & AD~v0 & AD~v1 & AD v2\\ \midrule

4.27$\pm$.10 & \multicolumn{2}{c}{3.28$\pm$.13} & 3.72$\pm$.12   & 3.90$\pm$.11   & \textbf{3.92}$\pm$.10   & 3.78$\pm$.12 \\
\bottomrule
\end{tabularx}%
}
\end{table}

\subsection{Subjective Evaluation}
To evaluate the perceptual quality of the codecs, we conducted MOS tests of a testing subset including randomly selected 15 utterances of each speaker. Baseline SS-, symAD-, and three vocoder-based AD-generated utterances were evaluated for their overall quality. Natural speech was also included as a reference, so the total number of testing utterances was 180. Ten subjects, either audio experts or native speakers, with headphones participated in the tests. As shown in Table~\ref{tb:subjective}, although there is still a gap between these coding and natural speech, the proposed AD-series codecs significantly outperform the baseline SS, which shows the effectiveness of the proposed mechanisms. The results also show that if the modeling capacity of a vocoder is advanced (e.g. AD~v1), the vocoder-based approaches achieve the best performance. Moreover, the competitive scores of AD~v1 and AD~v0 indicate that the ad~hoc kernel size is unnecessary, and the group convolutions work as well as MRF. These comparisons can be found on our demo page\footnote{\url{https://bigpon.github.io/AudioDec\_demo/}}.

\begin{table}[t]
\caption{Training speed w/ GPU A100 }
\label{tb:training}
\fontsize{8pt}{9.6pt}
\selectfont
{%
\begin{tabularx}{\columnwidth}{@{}p{1.7cm}YYYY@{}}
\toprule
& Encoder  & Decoder & Discriminator & Speed\\ \midrule
symAD stage1  & \checkmark & \checkmark &  $\times$ & 15.19~it/s \\
symAD stage2 & $\times$ & \checkmark &  \checkmark & 3.77~it/s \\
symAD* & \checkmark & \checkmark &  \checkmark & 3.43~it/s \\
\bottomrule
\end{tabularx}%
}
\end{table}

\subsection{Discussion}
To show the training efficiency of the proposed training paradigm, the symmetric AudioDec was adopted to conduct the training speed evaluations. The models were trained using one NVIDIA A100 SXM 80GB, and the training speed is presented by the average number of iterations in one second. The results in Table~\ref{tb:training} show that training with only metric loss is much fast than training with discriminators, so the proposed paradigm is effective for developing encoders of different applications. Moreover, fixing the encoder during the decoder and discriminator training also slightly improves the training speed. In conclusion, with the proposed training paradigm and an A100 GPU, training an encoder for a new application such as denoising takes only 3.5~hrs, which markedly improves the efficiency of new codec developments for different scenarios.

To evaluate the streaming capability, the processing times of segmental encoding and decoding with a nonoverlapping sliding window using the codecs were recorded. Although the overall latency is determined by the window length or the overall processing time, the streaming capability depends on only the longest processing time because the encoding and decoding can run in parallel. The evaluations were conducted using 50 randomly selected testing utterances on one NVIDIA GeForce RTX3090 GPU or AMD Ryzen Threadripper 3970X 32-core processor 3.70~GHz CPU with four threads.

As shown in Table~\ref{tb:latency_gpu}, because of the powerful GPU parallel computation capacity, the processing times of different window lengths are almost the same. We can find that the symAD, AD~v1, and AD~v2 codecs are potentially streamable (In a real scenario, there are some additional delays such as transmissions) even with 12.5~ms buffers, which demonstrates the effectiveness of the group convolutions to take advantage of the parallel computations for processing time reductions. On the other hand, even on the CPU with four threads as shown in Table~\ref{tb:latency_cpu}, the symAD and AD~v2 codecs are still potentially streamable with 12.5~ms buffers. According to our preliminary experiments, a stand-alone audio recording, encoding, decoding, and playing pipeline can work smoothly with a 25~ms buffer size on the GPU using AD~v1 and with a 35~ms buffer size on the CPU using AD~v2. Since the acceptable maximum latency of normal internet calls is 150~ms, there is still room for other processes. The streaming demo with the pretrained models is also released on our repository\textsuperscript{\ref{repo}}.

\begin{table}[t]
\caption{Latency analysis w/ GPU RTX3090 (ms) }
\label{tb:latency_gpu}
\fontsize{8pt}{9.6pt}
\selectfont
{%
\begin{tabularx}{\columnwidth}{@{}p{1cm}YYYYY@{}}
\toprule
\multirow{2}{*}{\begin{tabular}[c]{@{}l@{}}Window\\ length\end{tabular}} 
        & Encoder  & \multicolumn{4}{c}{Decoder}\\ \cmidrule(lr){2-2} \cmidrule(lr){3-6}
        & AD & sym & v0 & v1 & v2 \\ \midrule
12.5~ms & 4.8$\pm$.00 & 3.0$\pm$.00 & 12.7$\pm$.01 & 5.6$\pm$.00 & 5.4$\pm$.00 \\
25~ms & 6.0$\pm$.05 & 3.8$\pm$.04 & 13.1$\pm$.02 & 5.8$\pm$.01 & 5.5$\pm$.01 \\
50~ms & 5.2$\pm$.02 & 3.3$\pm$.01 & 14.0$\pm$.03 & 6.7$\pm$.02 & 6.2$\pm$.02 \\
100~ms & 5.1$\pm$.01 & 3.2$\pm$.00 & 13.2$\pm$.02 & 6.0$\pm$.03 & 5.7$\pm$.03 \\
\bottomrule
\end{tabularx}%
}
\end{table}

\begin{table}[t]
\caption{Latency analysis w/ CPU 3970X and 4 threads (ms)}
\label{tb:latency_cpu}
\fontsize{8pt}{9.6pt}
\selectfont
{%
\begin{tabularx}{\columnwidth}{@{}p{1cm}YYYYY@{}}
\toprule
\multirow{2}{*}{\begin{tabular}[c]{@{}l@{}}Window\\ length\end{tabular}} 
        & Encoder  & \multicolumn{4}{c}{Decoder}\\ \cmidrule(lr){2-2} \cmidrule(lr){3-6}
        & AD & sym & v0 & v1 & v2 \\ \midrule
12.5~ms & 6.8$\pm$.02 & 6.8$\pm$.01 & 28.7$\pm$.05 & 18.5$\pm$.02 & 9.5$\pm$.01 \\
25~ms & 8.2$\pm$.02 & 8.6$\pm$.03 & 35.1$\pm$.11 & 22.8$\pm$.07 & 11.2$\pm$.02 \\
50~ms & 9.0$\pm$.03 & 9.4$\pm$.03 & 37.6$\pm$.13 & 29.4$\pm$.12 & 14.1$\pm$.05 \\
100~ms & 11.8$\pm$.04 & 13.1$\pm$.06 & 45.1$\pm$.15 & 45.5$\pm$.30 & 21.2$\pm$.12 \\
\bottomrule
\end{tabularx}%
}
\end{table}

\section{Conclusion}
\label{sec:conclusion}
\vspace{-\baselineskip}
In this paper, we present an open-source neural audio codec, AudioDec, for high-fidelity 48~kHz audio. The proposed training paradigm markedly reduces the training time for a new encoder while achieving better quality. The proposed modularized architecture enables us to greatly improve the codec speech quality by using a powerful vocoder and advances the flexibility of developing codecs for different scenarios. The proposed low-latency implementations make AudioDec streamable in real-time with a 25~ms window length on both GPU and CPU. In conclusion, AudioDec is a high-quality, efficient, and convenient benchmark for audio codec research.



\vfill\pagebreak

\bibliographystyle{IEEEbib}
\bibliography{refs}

\end{document}